\def\lesssim{{_ <\atop{^\sim}}}

\def\ap3m{AP$^3$M}
\def\LCDM{$\Lambda$CDM}

\def\hkpc{$h^{-1}{\ }{\rm kpc}$}
\def\hMpc{$h^{-1}{\ }{\rm Mpc}$}
\def\Mpch{$h{\ }{\rm Mpc}^{-1}{\ }$}
\def\hMsun{$h^{-1}{\ }{\rm M_{\odot}}$}

\def\nbody{$N$-body}
\def\c15{$c_{\rm 1/5}$}

\newcommand{\Eq}[1]{Eq.~(\ref{#1})}
\newcommand{\Sec}[1]{Section~\ref{#1}}
\newcommand{\Fig}[1]{Fig.~\ref{#1}}
\newcommand{\Table}[1]{Table~\ref{#1}}
\newcommand{\mlapm}{\texttt{MLAPM}}

\def\ea{et~al.~}                            

\def\lesssim{\mathrel{\hbox{\rlap{\hbox{\lower4pt\hbox{$\sim$}}}\hbox{$<$}}}}
\def\gtrsim{\mathrel{\hbox{\rlap{\hbox{\lower4pt\hbox{$\sim$}}}\hbox{$>$}}}}

\newcommand{\AaA}[3]    {\mbox{A\&A~\textbf{#1},~#2~(#3)}}

\newcommand{\ApJ}[3]    {\mbox{ApJ~\textbf{#1},~#2~(#3)}}
\newcommand{\ApJS}[3]   {\mbox{ApJ~Suppl.~\textbf{#1},~#2~(#3)}}
\newcommand{\ApJL}[3]   {\mbox{ApJ~Lett.~\textbf{#1},~#2~(#3)}}

\newcommand{\MNRAS}[3]  {\mbox{MNRAS~\textbf{#1},~#2~(#3)}}

\newcommand{\PhRevD}[3] {\mbox{Phys.~Rev.~\textbf{D#1},~#2~(#3)}}
\newcommand{\astroph}[1]{\mbox{\texttt{astro-ph/#1}}}

\documentstyle[epsfig,12pt]{article}
\def\reference{\parskip 0pt\par\noindent\hangindent 0.5 truecm}
\begin{document}
\title{On the reliability of initial conditions for dissipationless 
       cosmological simulations}
%

\author{Alexander Knebe $^{1}$, Alvaro Dom\'\i nguez$^2$
} 

\date{}
\maketitle

{\center
$^1$ Centre for Astrophysics \& Supercomputing,
     Swinburne University, P.O. Box 218, Mail \# 31,
     Hawthorn, Victoria, 3122, Australia\\
$^2$ Max-Planck-Institute f\"ur Metallforschung, 
     Heisenbergstr.~3, 70569 Stuttgart, Germany\\[3mm]
}

\begin{abstract}

We present the study of ten random realizations of a density field
characterized by a cosmological power spectrum $P(k)$ at redshift
$z=50$. The reliability of such initial conditions for \nbody\
simulations are tested with respect to their correlation
properties. 
The power spectrum $P(k)$, and the mass variance $\sigma_M(r)$ do not
show detectable deviations from the desired behavior in the
intermediate range of scales between the mean interparticle distance
and the simulation volume. The estimator for $\xi(r)$ is too noisy to
detect any reliable signal at the initial redshift $z=50$.
The particle distributions
are then evolved forward until
$z=0$. This allows us to explore the cosmic variance stemming from the
random nature of the initial conditions. 
With cosmic variance we mean the fact that a simulation represents a
single realization of the stochastic initial conditions whereas the real
Universe contains many realizations of regions of the size of the box; this problem
affects most importantly the scales at about the fundamental mode. 
We study morphological descriptors of the matter distribution such as
the genus, as well as the internal properties of the largest object(s)
forming in the box. We find that the scatter is at least comparable to
the scatter in the fundamental mode.

\end{abstract}

{\bf Keywords: methods: n-body simulations -- cosmology: dark matter}

\bigskip

\section{Introduction} \label{Intro}

Our present understanding of the formation and properties of the
cosmological large-scale structure relies to a large extent on \nbody\
simulations: given the difficulty in addressing theoretically the
highly nonlinear regime of the growth of density inhomogeneities by
the gravitational instability, simulations have proven a valuable tool
to get insight into the (non-linear) structure formation
scenarios. Therefore, it is of considerable importance to confirm the
reliability of such simulations. 

It has been claimed recently (Baertschiger~\& Sylos Labini 2002) that
there are major problems with generating initial conditions (ICs) for
\nbody\ simulations. 
We can identify several reasons why the ICs {\em may} introduce
uncertainties in the subsequent evolution. First, there is the problem
of finite-mass resolution or {\em discreteness}: the initial continuum density
field is modeled by the distribution of a {\em finite} number of point
particles $N$, therefore only a finite number of Fourier modes of the
density field can be reproduced reliably. The maximum wavenumber ({\it
Nyquist wavenumber}) is given by $k_{\rm max} = \pi/\Delta x$, where
$\Delta x$ is the mean interparticle separation.  The modes $k
\gtrsim k_{max}$ have spurious values related to the point-particle
distribution and may lead to artificial effects in the posterior
dynamical evolution.  The finite-mass resolution is expected to be
irrelevant if the nonlinear mode-mode coupling to the modes $k \gtrsim
k_{max}$ has only a small influence on the dynamics.

The second problem with the ICs is the finite size of the simulation
box with side length $B$, which implies that the values of the Fourier
modes of the density field with wavenumber smaller than the {\it
fundamental wavenumber}, $k < k_{\rm min} = 2\pi/B$, are artificially
set to zero. This leads to two possible difficulties: first, the
absence of mode-mode coupling to those large-scale modes, and second
the so-called {\em cosmic variance}, meaning that the simulation box
represents only one (finite-sized) realization of the stochastic
initial density field, whereas the true Universe contains many
realizations of regions of the size of the box. 
Therefore, the morphological properties of the matter distribution in
a certain volume, as measured by e.g. the genus statistics, will
presumably show some intrinsic scatter when placing the volume at
different locations in the real Universe. And this will also happen
with the (internal) properties of any given class of objects. This is
one of the main aspects of the current
study and what we refer to as cosmic variance (in \nbody\ simulations)
throughout the paper even though one might argue that this is not the
"real" cosmic variance but rather an artificially introduced
\textit{sampling variance}. However, we are actually interested in
exactly that (sampling) effect which can easily be tested by simply
running the same cosmological simulation but using different random
realizations of the initial density field.

In this work we study systematically the reliability of the initial
density field used as an input to the \nbody\ simulations as well as
the effect of their random nature onto the internal properties of
clusters. In \Sec{nbody} we briefly review the most commonly way used
to generate ICs and the code used to evolve the particles into the
non-linear regime. \Sec{DManalysis} focuses on some of the statistical
characteristics of the Dark Matter field: we consider the 2-point
correlation function, the power spectrum, the mass variance in
spheres, and the Minkowski functionals, the latter being sensitive to correlations of higher order.  
Finally, in \Sec{HaloAnalysis} we investigate dark matter
clusters identified within the simulations and quantify the effect of
the cosmic variance on their internal properties.

\section{\nbody\ Simulations} \label{nbody}

\subsection{Generating Initial Conditions} \label{ICs}
The commonly used way for setting up initial conditions for a
cosmological simulation is to make use of the Zeldovich approximation
to move particles from a Lagrangian point $\vec{q}$ to an Eulerian point
$\vec{x}$ (e.g. Efstathiou, Frenk~\& White 1985):

\begin{equation} \label{zeldovich}
 \vec{x} = \vec{q} - D(t) \vec{S}(\vec{q}) \ ,
\end{equation}

\noindent
where $D(t)$ describes the growing mode of linear fluctuations and
$\vec{S}(\vec{q})$ is the 'displacement field'. This method is not
restricted to a cosmological scenario nor to the Zeldovich
approximation: it is very general, relying only on the continuity
equation for the transport of particles in the limit $D(t)
\rightarrow 0$. The initial Lagrangian coordinates $\vec{q}$ are
usually chosen to form a regular, three-dimensional lattice although
there are other possible point-particle realizations yielding a
homogeneous and isotropic density field on large scales (i.e.
glass-like initial conditions, White 1996).

For the runs presented in this study we used the code described in
Klypin~\& Holtzmann (1997) to set up the initial conditions

\begin{equation} \label{FourierSum}
 \vec{S}(\vec{q}) = \nabla_q \Phi(\vec{q}), \ \ \
 \Phi(\vec{q})    = \sum_{\vec{k}} a_{\vec{k}} \cos(\vec{k} \cdot \vec{q}) + 
                                   b_{\vec{k}} \sin(\vec{k} \cdot \vec{q}) \ ,
\end{equation}

\noindent
where the Fourier coefficients $a_{\vec{k}}$ and $b_{\vec{k}}$ are related to 
a pre-calculated input power spectrum of density fluctuations, $P(k)$, as follows:

\begin{equation} \label{FourierAmplitudes}
 a_{\vec{k}} = R_1 \frac{1}{k^2} \sqrt{P(k)} , \ \ \ 
 b_{\vec{k}} = R_2 \frac{1}{k^2} \sqrt{P(k)} .
\end{equation}

\noindent
$R_1$, $R_2$ are (Gaussian) random numbers with mean zero and dispersion
unity. The factor $1/k^2$ is (the Fourier transform of) the Green's
function of Poisson's equation\footnote{Actually, $-1/k^2$ is the
correct Green's function, but the factor $-1$ can be dropped as $R_1$
and $R_2$ scatter around zero.} and $\Phi(\vec{q})$ can therefore be understood
as the gravitational potential created by a Gaussian stochastic
density field whose power spectrum agrees with the input $P(k)$; the
power spectrum $P(k)$ measures the strength of each individual
$k$-mode contributing to the density field.  However, to fully
preserve the random nature both amplitudes (sine- and cosine-wave) are
to be picked from a Gaussian distribution.

\Eq{FourierSum} can be rewritten introducing complex numbers:

\begin{equation} \label{Pkcomplex}
 \Phi(\vec{q}) = \sum_{\vec{k}} 
                 A_{\vec{k}} \exp(\imath [\vec{k} \cdot \vec{q} + \theta_{\vec{k}}]) , \quad 
                 A_{\vec{k}} \exp(\imath \theta_{\vec{k}}) := \frac{1}{2} [a_{\vec{k}} + a_{-\vec{k}} - 
                 \imath (b_{\vec{k}} - b_{-\vec{k}})].
\end{equation}



Both $A_{\vec{k}}$ and $\theta_{\vec{k}}$ need to be drawn from
appropriate random distributions. However, the ICs of cosmological
relevance are ergodic for $A_{\vec{k}}$ with $k \gg k_{min}$, making
their random nature irrelevant: spatial regions of size much smaller
than the simulation box already work as many different realizations
inside the box for those amplitudes. Thus, cosmic variance enters
through the random nature of the phases $\theta_{\vec{k}}$ and of the
amplitudes $A_{\vec{k}}$ for $k \approx k_{min}$.

The idea of this paper is to create a certain number of random
realizations of the same power spectrum $P(k)$ by using different
random seeds when drawing $R_1$, $R_2$ in \Eq{FourierAmplitudes}. The input
power spectrum $P(k)$ was calculated using the CMBFAST code (Seljak~\&
Zaldarriaga 1996), and all parameters (e.g. box size, number of
particles, force resolution, integration steps, etc.) were fixed
except for the seed for generating the random sequence providing the
$R$--values\footnote{An appropriate routine might be
\texttt{GASDEV} from Numerical Recipes (Press~\ea 1992).}.

\subsection{Simulation Details}
We created a data base of ten simulations that all were started at a
redshift $z_i=50$ and evolved until $z=0$ in a \LCDM\ ($\Omega_0 =
0.3, \Omega_\lambda = 0.7, \Omega_b h^2 = 0.04, h = 0.7, \sigma_8 =
0.9$) cosmological model using 128$^3$ particles within a box of side
length $B$ = 64\hMpc, giving a mass resolution of $m_p = 1.04\cdot
10^{10}$\hMsun. They were carried out using the publicly available
adaptive mesh refinement code \mlapm\ (Knebe, Green~\& Binney 2001).
\mlapm\ reaches high force resolution by refining all high-density 
regions with an automated refinement algorithm.  The refinements are
recursive: the refined regions can also be refined, each subsequent
refinement having cells that are half the size of the cells in the
previous level.  This creates an hierarchy of refinement meshes of
different resolutions covering regions of interest.  The refinement is
done cell-by-cell (individual cells can be refined or de-refined) and
meshes are not constrained to have a rectangular (or any other)
shape. The criterion for (de-)refining a cell is simply the
number of particles within that cell and a detailed study of the
appropriate choice for this number can be found elsewhere (Knebe,
Green~\& Binney, 2001). The code also uses multiple time steps on
different refinement levels where the time step for each
refinement level is two times smaller than the step on the
previous level. A regular 256$^3$ domain
grid was used to cover the whole computational volume in all runs
and cells were
refined as soon as the number of particles per cell exceeded the
preselected value of 8. We stored snapshots of the particle
distribution at redshifts $z=5$, $z=1$, $z=0.5$, $z=0.25$, and $z=0$.
At the end of the runs the force resolution is determined by the
highest refinement level reached: for the runs at hand the finest grid
at $z=0$ consisted of 8192 cells per side and was called into
existence at redshift $z \sim 0.88$. This grid corresponds to a force
resolution of about 23\hkpc\ which is simply three times the grid
spacing and gives the scale where the forces are purely
Newtonian. This is sufficient enough for the presented study as we are
mainly interested in the overall (large-scale) clustering properties.
But as we will see later on, we are resolving approximately 2\% of the
virial radius of the most massive halo formed in the runs, which is
sufficient for investigations of the internal properties such as
velocity dispersion, spin parameter and triaxiality.

\section{Analysis I: Dark Matter Field} \label{DManalysis}
We first focus on the properties of the dark matter particle
distributions. Our main aim is to assess the recent claims by
Baertschiger~\& Sylos Labini (2002) that there are major problems with
generating initial conditions for \nbody\ simulations in the way as
outlined in \Sec{ICs}. Because it is common to either use a regular
grid or a glass-like distribution as Lagrangian starting points
$\vec{q}$ for the Zeldovich approximation (cf. \Eq{zeldovich}), their
arguments try to prove that this leads to spurious artifacts related
to e.g. the regular structure of such a grid, and that the initial
conditions are not able to reflect the superposed CDM-like
fluctuations at all.

\begin{figure}
 \centerline{\psfig{file=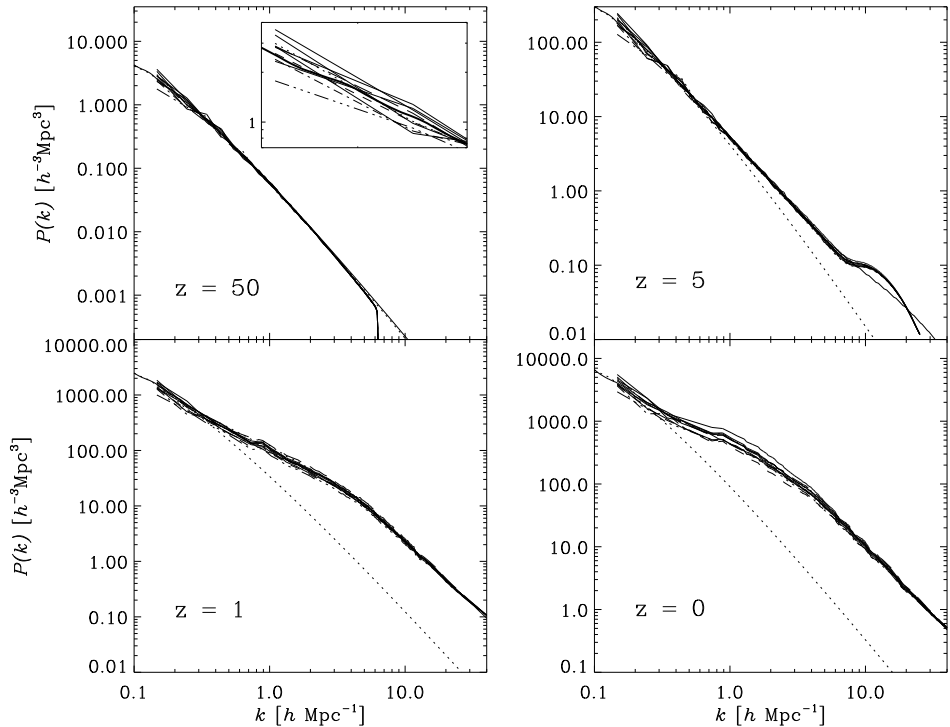,width=13cm}}
 \caption{Power spectrum evolution for all ten runs as compared to the
          prediction by Peacock~\& Dodds (1996) (solid line) and the linear $P(k)$ 
          (dotted line), respectively. The inset panel for $z=50$
          focuses on the fundamental mode $k_{\rm min}=2\pi/B$ which
          shows a 1$\sigma$ variance of approximately 20\%.}  
 \label{Power64xx} 
\end{figure}

\subsection{Power Spectra}
When creating a fluctuating density field in a certain volume by using
a fixed number of particles, one is limited in the range of $k$'s by
the size of that volume on the one hand, and the number of particles
used to sample the waves on the other hand. The wave number of the
lowest frequency wave (fundamental mode) to fit into the box is given
by $k_{\rm min} = 2\pi/B$ where $B$ is the side length of the box. The
maximum wave number is determined by the Nyquist frequency, $k_{\rm
max} = \pi / \Delta x$, where $\Delta x = B/N^{1/3}$ is the mean particle 
separation (not to be confused with the grid spacing used in the
\nbody\ code and for extracting the power spectrum from such a
particle distribution, respectively).
A recent investigation showed that high-resolution \nbody\ simulations
where even smaller scales than $k_{\rm max}$ are resolved are
justified for power spectra with an effective spectral index $n_{\rm
eff} = d \log P(k) / d \log k$ much less than -1 (Hamana, Yoshida~\&
Suto 2001). And this is the case for (nearly) all CDM type spectra as
$P(k) \propto k^{-3}$ for large $k$. The evolution of power on small
scales is driven by the transfer of power from large scales and hence
it is important to follow that evolution with an appropriate force
resolution even though that small-scale power was not present in the
initial conditions (see Introduction).

Using the particle data at redshifts $z=50$ (initial conditions),
$z=5$, $z=1$, and $z=0$, we derived $P(k)$ by Fourier transforming the
density field on a regular 512$^3$ grid\footnote{The density was
assigned to the grid cells using the Triangular-Shape-Cloud method.},
which effectively introduces $k_{\rm max} \approx 25$\Mpch\ as the
maximum wavenumber to be recovered from the data. We adopted the
method for extracting even higher $k$ waves from the particle
distributions as outlined in Jenkins~\ea (1998). The power spectra
were then compared to the non-linear prediction given by Peacock~\&
Dodds (1996, PD96). However, their fitting parameters depend on the
spectral slope $n=d\ln{P(k)}/d\ln{k}$ and hence some recipe needs to
be adopted when applying it to a cosmological $P(k)$ where $n$ is a
function of $k$. We used $n_{\rm eff}$ defined via

\begin{equation} \label{neff}
 n_{\rm eff}(k_l) = \left. \frac{d\ln{P(k)}}{d\ln{k}} \right|_{k=k_l/2}
\end{equation}

\noindent
for the estimate  of the spectral index $n_{\rm eff}$ at wave
number $k_l$ (cf. Jenkins~\ea 1998 and Peacock~\& Dodds 1996).

In \Fig{Power64xx} the results are shown along with the linearly
extrapolated $P(k)$. There are a couple of things to note besides the
overall good agreement of the estimated $P(k)$ with the PD96
prediction: first, the power spectrum derived from the particle
distribution agrees extremely well with the input $P(k)$\footnote{At
redshift $z=50$ the non-linear $P(k)$ as given by PD96 is
indistinguishable from the linear $P(k)$.} and the fundamental mode
has {\em not} turned non-linear at $z=0$; second, we can clearly see
how the scatter prominent in the large waves $k_{\rm min} \gtrsim
2\pi/B$ at high redshifts moves towards higher $k$-values at later
times. The scatter in the initial conditions is of the order of 20\%
and it arises because only a small number of such harmonics do fit
into the finite box of side length $B$. We ascribe the migration of
the scatter downwards to smaller scales (higher $k$-values) to the
transfer of power from large to small scales: the higher the amplitude
$A_{\vec{k}}$ at $k_{\rm min}$ (cf. \Eq{Pkcomplex}) the more power can be
transfered to smaller $k$'s. And hence we are facing a faster
evolution of small-scale structures leading to the observed dispersion
amongst the individual runs.

The discrepancy with the PD96 prediction for $z=5$ for $k>10$ is not
physical. Even though the \mlapm\ code invoked already three levels of
refinements at $z=5$ they are still very small in size, i.e. there are
only ca. 40,000 refinement cells in total with about 12,000 particles
($\sim$ 0.6\% of all particles) being moved on those levels. However,
a visual inspection of the refined regions shows that the grids are
covering all prospective halo formations sites and hence we are
following the built-up of structures correctly. But when trying to
recover those high-$k$ modes from the simulations we are left with
observed mismatch due to the majority of the particles still being
moved on the 256$^3$ grid.

\subsection{Mass Variance $\sigma(r)$}
The variance $\sigma_M$ of the mass in spheres with radius $r$ is given by

\begin{equation}\label{sigmarAna}
 \sigma_M^2(r) = \frac{1}{2\pi^2} \int_0^{+\infty} P(k) \hat{W}^2(kr) k^2 dk ,
 \qquad \hat{W}(x) = \frac{3}{x^3} (\sin{x} - x\cos{x}) \ .
\end{equation}

%

\noindent
The function $\sigma^2_M(r)$ is readily calculated and can be compared
to an adequate estimator $\sigma_{M, {\rm est}}^2(r)$ when being
applied to the actual particle data. Our estimator distributes a
certain number of spheres with radius~$r$ at random in the simulation
volume and compares the number of particles inside those spheres to
the expected mean value:

\begin{equation}\label{sigmarSimu}
 \sigma_{M, {\rm est}}^2(r) = \frac{1}{\langle N_r\rangle^2} \sum_{i=1}^{N_s} 
                              \frac{(N_i(r) \ - \langle N_r\rangle)^2}{N_s - 1} \ .
\end{equation}

\noindent
$N_s$ is the total number of spheres with radius~$r$ and 
$\langle N_r\rangle=\langle \rho\rangle4\pi r^3 / 3 m_{\rm p}$ is the mean number of 
particles in such a sphere.

\subsubsection{Reliability of Estimator}

   \begin{figure}
      \centerline{\psfig{file=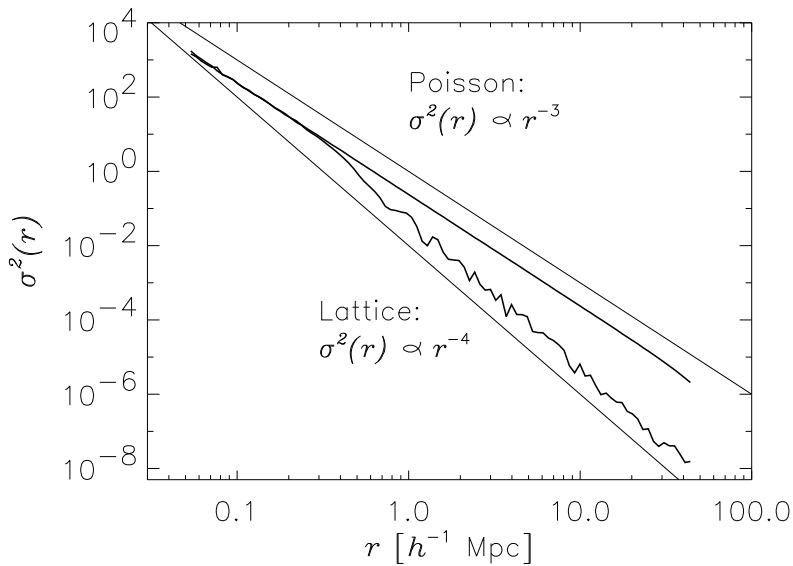,width=11cm}}
      \caption{Reliability check for our $\sigma^2_{M, \rm est}(r)$ 
               estimator \Eq{sigmarSimu}. 
               The solid lines have the slopes of the analytical 
               expectations (refer to the text for further details).
               All amplitudes are arbitrary.}
      \label{SigmaRestimator} 
   \end{figure}

In order to make sure our estimator works as expected we started by
applying it to particle distributions where simple scaling laws for
$\sigma_M^2(r)$ can be calculated analytically. 
For a purely Poissonian particle distribution one derives easily

\begin{equation} \label{sigmaPoisson}
 \sigma_{M, {\rm Poisson}}^2(r) \propto r^{-3} \ ,
\end{equation}

\noindent
and for a ``shuffled'' lattice (e.g. Gabrielli, Joyce~\& Sylos Labini
2002):

\begin{equation}
 \sigma_{M, {\rm Lattice}}^2(r) \propto r^{-4} , \qquad (r \gg \textrm{lattice spacing})\ .
\end{equation}

\noindent
From \Fig{SigmaRestimator} we deduce that our estimator does indeed
work correctly: we created ten Poisson distributions of 128$^3$
particles in a (128\hMpc)$^3$ volume and for each distribution we
calculated $\sigma_{M, {\rm est}}^2(r)$ using 10000 spheres (for each
$r$ value). The curve shown is the average taken over the ten Poisson
distributions. The error bars are too small to be presented. The
shuffled lattice distribution was created as follows: we placed
128$^3$ particles on the nodes of a 128$^3$ grid with spacing 1\hMpc,
and then each particle was shifted in $x$, $y$, and $z$ directions by
a random amount uniformly distributed in the range $[-0.05, 0.05]$
\hMpc; ten such realizations were created. The curve shown in
\Fig{SigmaRestimator} is again the mean estimate when averaging over
the ten realizations. In both tests we recover the expected scaling
relation.

\subsubsection{Application to \nbody\ Data}

   \begin{figure}
      \centerline{\psfig{file=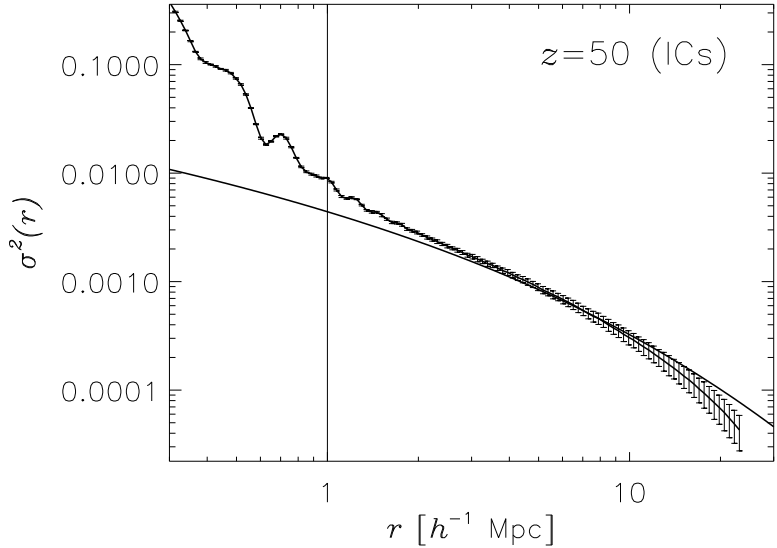,width=11cm}}
      \caption{Mean value of $\sigma_M^2(r)$ as given 
               by \Eq{sigmarSimu} when averaged over the ten initial
               conditions at redshift $z=50$. 
               The error bars are 1$\sigma$ errors. The solid line 
               is the analytical expectation
               given by \Eq{sigmarAna}.
               The vertical line indicates the scale corresponding
               to the (particle) Nyquist frequency, $k_{max}$.}
      \label{SigmaRic} 
   \end{figure}

We now apply the estimator \Eq{sigmarSimu} to our ICs
as well as the final outputs at redshift $z=0$. \Fig{SigmaRic} shows
$\sigma^2_{M, \rm est}(r)$ compared to the analytical $\sigma^2_M(r)$
as given by \Eq{sigmarAna}.  For every scale~$r$ we used again $N_s =
10000$ randomly placed spheres.  The mean mass variance $\langle
\sigma_{M, {\rm est}}^2(r)\rangle_{\rm set}$ (averaged over the ten
realizations) is plotted and the error bars are 1~times the variance
of $\sigma_{M, {\rm est}}^2(r)$ around the mean value
\mbox{$\langle \sigma_{M, {\rm est}}^2(r)\rangle_{\rm set}$}.

Contrary to the findings of Baertschiger~\& Sylos Labini (2002), we
observe that the initial conditions agree from approximately the scale
of the particle Nyquist frequency out to nearly half the box size with
the analytical predictions. The
faster drop of $\langle \sigma_{M, {\rm est}}^2(r)\rangle_{\rm set}$
for scales approaching the box size is simply the effect of the finite
(periodical) box. As soon as the volume of the sphere comes close to
the actual box size (which happens for $r \approx B/2$) one finds
nearly all particles in the sphere due to the periodic boundary
conditions. Hence the variance $\sigma_{M, {\rm est}}^2(r)$ drops
faster than predicted by \Eq{sigmarAna}. And the larger amplitude of
$\sigma_{M, \rm est}^2(r)$ for small scales is indeed a reflection of
the discreteness of the initial conditions. But in any case
\Fig{SigmaRic} is a rather convincing argument that the mass variance 
in the initial conditions does agree with the CDM type fluctuations as
described by the input power spectrum $P(k)$.

   \begin{figure}
      \centerline{\psfig{file=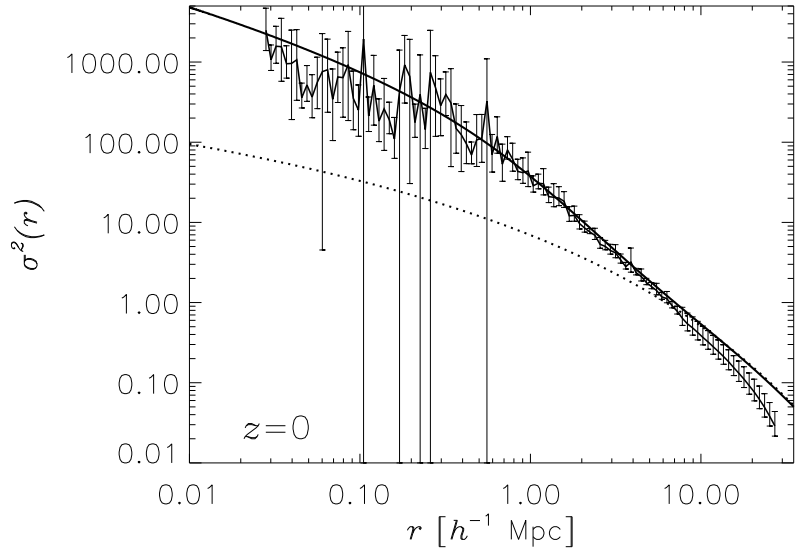,width=11cm}}
      \caption{Same as \Fig{SigmaRic} but this time for the final
               output at $z=0$. The two analytical curves are the 
               linear extrapolation of \Eq{sigmarAna} to $z=0$
               (dotted line) and the prediction for $\sigma^2(r)$
               when using the PD96 power spectrum with \Eq{sigmarAna}
               (solid line).}
      \label{SigmaRz0} 
   \end{figure}

When comparing $\sigma^2_{M, \rm est}(r)$ for the final output at
redshift $z=0$ to the analytical $\sigma_M^2(r)$ in \Fig{SigmaRz0}
(both using the linearly extrapolated $P(k)$ as well as the non-linear
$P(k)$ given by PD96) we notice again a couple of things: firstly,
the large scales ($r \gtrsim B/2$) are still below the expectation,
and secondly, there is more pronounced scatter on scales $r <
0.6$\hMpc\ than found for the ICs.  For $r \gtrsim B/2$, the explanation is 
again the finite periodic box. The increased value for the variance 
$\sigma^2_{M, \rm
est}(r)$ for small scales $r < 1$\hMpc\ (and its large scatter) is
readily explained by the fact that gravitationally bound objects (and 
voids) are forming which introduces some sort of
"semi-discreteness": this gives rise to a higher variance (as
well as larger scatter) on scales related to the average size of
such objects, i.e. $\sim$1\hMpc\ and below.

\subsection{Two-Point Correlation Function $\xi(r)$}

The two-point correlation function is the Fourier transform of the
power spectrum:

\begin{equation} \label{xirAna}
 \xi(r) = \frac{1}{2\pi^2} \int_0^{+\infty} P(k) \frac{\sin(k r)}{k r} k^2 dk \ .
\end{equation}

\noindent
The basic interpretation of $\xi(r)$ is that it is the average number
of neighbors to a given object with distance $r$ in excess to a
Poisson distribution. And this is how we realize an
estimator for $\xi(r)$.  We start again putting down a certain number
of spheres in the simulation box, but this time centered at
particles. We then create a shell of thickness $dr$ extending from $r$
to $r+dr$. The correlation function can now be estimated by

\begin{equation} \label{xirSimu}
 \xi_{\rm est}(r) = \frac{\Gamma(r,dr)}{\langle \rho\rangle} - 1 \ ,
\end{equation}

\noindent
where $\langle \rho\rangle$ is the mean number density of the simulation and
$\Gamma(r,dr)$ the mean number density of particles found in the
shell $[r,r+dr]$:

\begin{equation}
\label{Gammai}
\Gamma(r,dr) = \frac{1}{N_s} \sum_{i=1}^{N_s} \Gamma_i(r,dr) \ .
\end{equation}

\noindent
For each value of $r$, we use again $N_s=10000$ shells $[r,r+dr]$ centered at a randomly
chosen particle for calculating the average number density of
particles $\Gamma(r,dr)$ .

\subsubsection{Reliability of Estimator}
This time it is more difficult to calibrate the estimator \Eq{xirSimu}
because our test models (Poisson and shuffled lattice) consist of Dirac $\delta$'s and zones of 
vanishing correlation.

\begin{figure}
      \centerline{\psfig{file=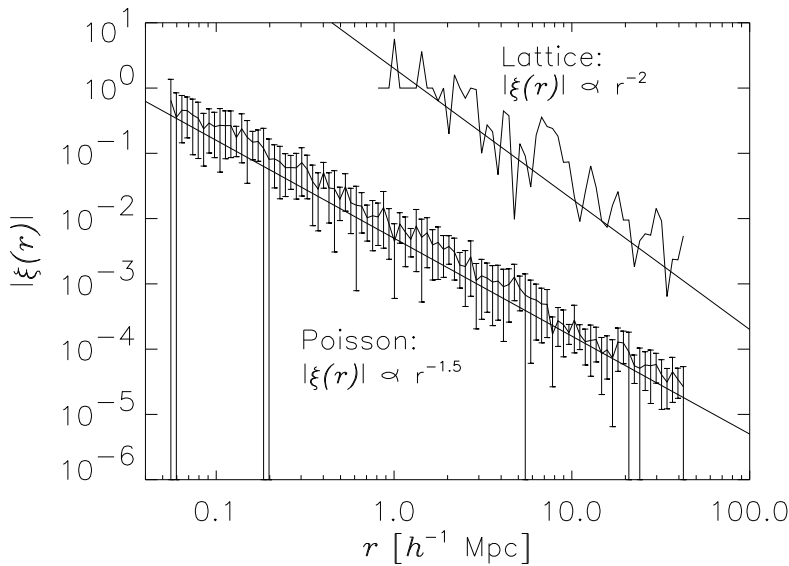,width=11cm}}
      \caption{Scaling relations for $|\xi_{\rm est}(r)|$ when
               applied to a set of Poisson distributions and of shuffled
               lattices with spacing 1\hMpc. The error
               bars for the Poisson distribution are 1$\sigma$ and
               for the lattice too small to be presented. The amplitudes
               are again arbitrary.}
      \label{XiRestimator} 
   \end{figure}

For the Poissonian case, $\xi(r)=0$ if $r \neq 0$, so that we expect
$\xi_{est} (r)$ to fluctuate around zero with an amplitude
proportional to the dispersion of the estimator, $\langle \xi^2_{est}
(r) \rangle$. Absence of correlations allows an easy estimation of the
dispersion: if $N_s$ is not too large (so that the probability that shells overlap is small), the numbers $\Gamma_i$ in Eq.~(\ref{Gammai}) are
uncorrelated of each other and have a Poissonian distribution. One can
then show immediately:
\begin{equation}
  \langle \Gamma_i \rangle = \langle \varrho \rangle , \qquad 
  \langle \Gamma_i \Gamma_j \rangle - \langle \Gamma_i \rangle \langle \Gamma_j \rangle = 
  \frac{\langle \Gamma_i \rangle}{4 \pi r^2 d r}   \delta_{ij} ,
\end{equation}
and then
\begin{equation}
  \langle \xi^2_{\rm est} (r) \rangle = \frac{1}{N_s \langle \varrho \rangle 4 \pi r^2 d r}.
\end{equation}
We took $d r \propto r$ (logarithmic binning), so that we expect the
amplitude of the fluctuations in $\xi_{est} (r)$ to decay like
$r^{-3/2}$, as indeed observed in \Fig{XiRestimator}, where the error
bars are again 1$\sigma$ errors when averaging over the ten random
sets. The Figure also shows $|\xi_{\rm est}(r)|$ for the shuffled
lattice with grid spacing of 1\hMpc. We believe that the observed
$r^{-2}$-decay is again due to $\langle \xi^2_{est} (r) \rangle$, like
in the Poissonian case.

\subsubsection{Application to \nbody\ Data}

\begin{figure}
      \centerline{\psfig{file=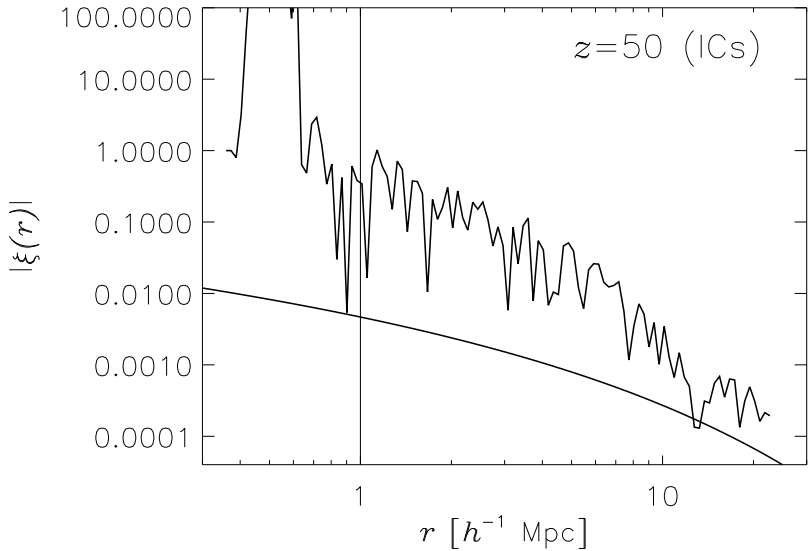,width=11cm}}
      \caption{Two-point correlation function for initial
               particle distribution at redshift $z=50$.
               For clarity no error bars are shown due to a 
               high level of noise. The solid line
               is the expected $\xi(r)$ as given by \Eq{xirAna}.
               We plot $|\xi(r)|$, as the estimated function
               tends to oscillate around zero for scales
               smaller than the Nyquist wavelength (indicated
               by the vertical line).}
      \label{XiRic} 
   \end{figure}

\Fig{XiRic} shows the result of applying the estimator \Eq{xirSimu} to the actual \nbody\ data 
at the initial redshift $z=50$. We plot the absolute value $|\xi(r)|$
as the correlation function tends to oscillate around zero, too.
The curve is the average
taken over the ten runs (as usual), but we do not plot the error bars
as the data already show a noticeable level of noise.  
%
This noise is in fact so strong as to mask the signal (the \LCDM\ 
behavior in this case); we already found this problem with the test
models especially for the lattice distribution upon which the
ICs are based on (cf. \Eq{zeldovich}). It seems that an improvement of the estimator (\ref{xirSimu})
is required to extract reliable information in these extreme cases.

\begin{figure}
      \centerline{\psfig{file=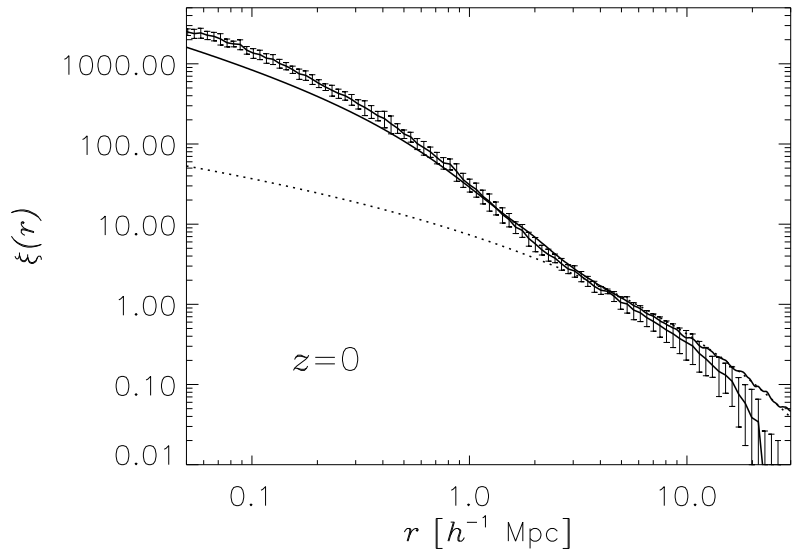,width=11cm}}
      \caption{Same as \Fig{XiRic} but for $z=0$. The error bars
               are 1$\sigma$ errors when averaging over the ten runs.
               The two lines are again the analytical PD96 fit (solid) 
               and the linear theory (dotted).}
      \label{XiRz0} 
   \end{figure}

\Fig{XiRz0} shows the same quantity for the $z=0$ data, where
the analytical curves are the correlations derived from the linearly
extrapolated $P(k)$ (dotted line) and the non-linear $P(k)$ 
from PD96 (solid line) in \Eq{xirAna}, respectively. This time we 
find a deviation of the estimated $\xi(r)$ from the one predicted
using PD96 on small scales. However, the error bars are 'only'
1$\sigma$ and the PD96 prediction still lies within the 3$\sigma$
level. 
One must also note that the estimator \Eq{xirSimu} is biased towards high
density regions where most of the particles at $z=0$ will reside,
since the shells are centered at (randomly chosen) particle positions
rather than placing them randomly at any point (like the estimator
\Eq{sigmarSimu} does, which also explains why the scatter for
$\xi^2_{\rm est}(r)$ at these small scales is much smaller than for
$\sigma^2_{M, \rm est}(r)$ observed in \Fig{SigmaRz0}). However, we
varied the number of spheres $N_s$ from 50 to 100000 and could only
detect a mild (if any) dependence of the amplitude on $N_s$.

Nevertheless, if one is to believe this discrepancy, then it is not
obvious which one is to be blamed, the simulations or the PD96 fit.
We have confirmed the excellent agreement of PD96 with our estimated
$P(k)$ in the probed range of wavenumbers (see \Fig{Power64xx}).
However, when using \Eq{xirAna}, one is extrapolating the PD96 fit to
all wavenumbers.  Clearly, the discrepancy should originate from the
modes beyond the probed range, but with the information at hand, one
cannot conclude whether their effect is estimated wrongly by the
simulation or by the PD96 fit.

\subsection{Minkowski Functionals}

We have computed also the four scalar Minkowksi functionals (MF) of
each realization (Mecke, Buchert~\& Wagner 1994). The MFs are
morphological measures of the structure, sensitive to correlations of
order higher than the second.  They include the {\it genus} statistics
(Melott 1990) and have been used to quantify how filamentary or
sponge-like the matter distribution looks like (Schmalzing \ea 1999),
to study galaxy distribution in catalogs (Kerscher \ea 1997), and to
address Gaussianity in the cosmic microwave background (Schmalzing~\&
Gorski 1998).

Like in deriving the power spectrum, we constructed a density field on
a regular grid using the Triangular-Shape-Cloud method with two
different resolutions: a $128^3$-cell grid and a $32^3$-cell grid. A
density threshold was introduced and the boundary surface was
determined between regions with a density below the threshold and
regions with a density above it. Finally, the MFs of the boundary
surface were determined. The four MFs are defined as follows:
\begin{itemize}
\item $M_0 = \mbox{}$ volume enclosed by the surface,
\item $M_1 = \mbox{}$ area of the surface,
\item $M_2 = \mbox{}$ integral over the surface of its mean curvature,
\item $M_3 = \mbox{}$ integral over the surface of its Gaussian
  curvature, which coincides with the Euler characteristic (genus):
  \begin{displaymath}
    M_3 = \textrm{number of disconnected objects + number of holes $-$ number of tunnels}.
  \end{displaymath}
\end{itemize}

\begin{figure}
  \centerline{\psfig{file=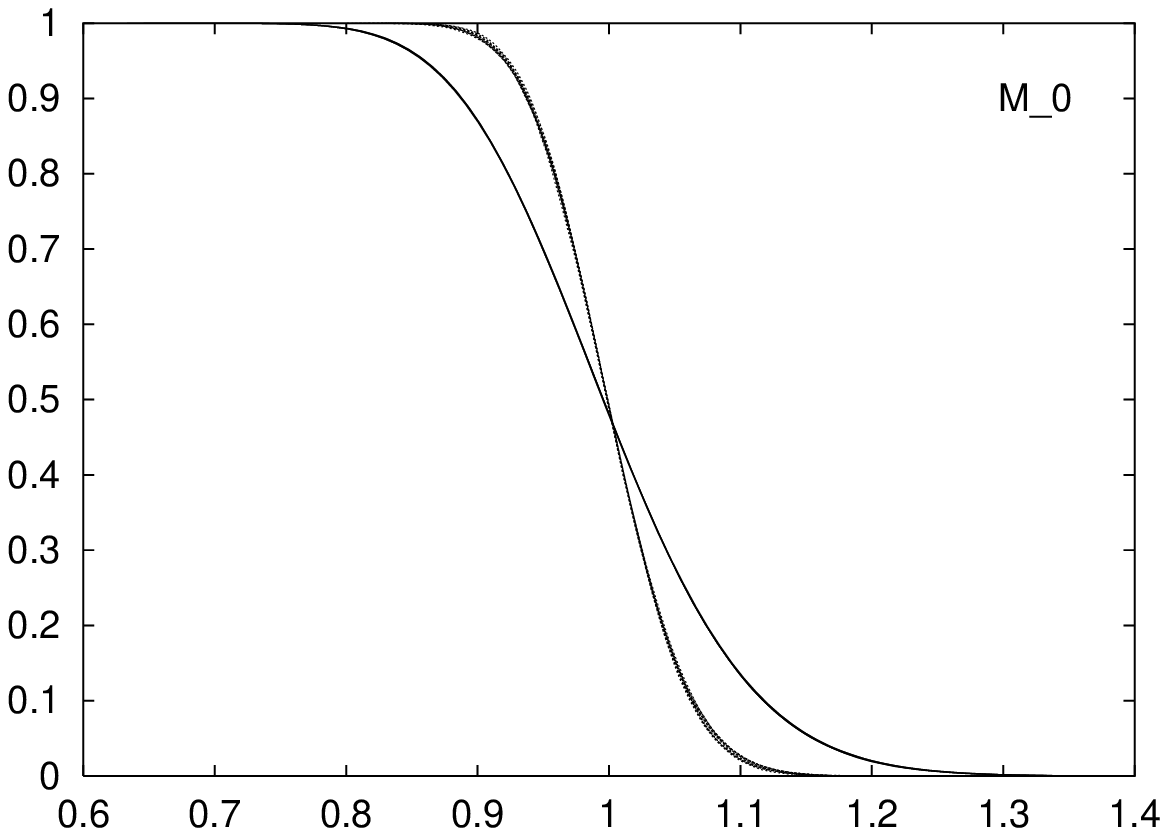,width=.4\hsize}\psfig{file=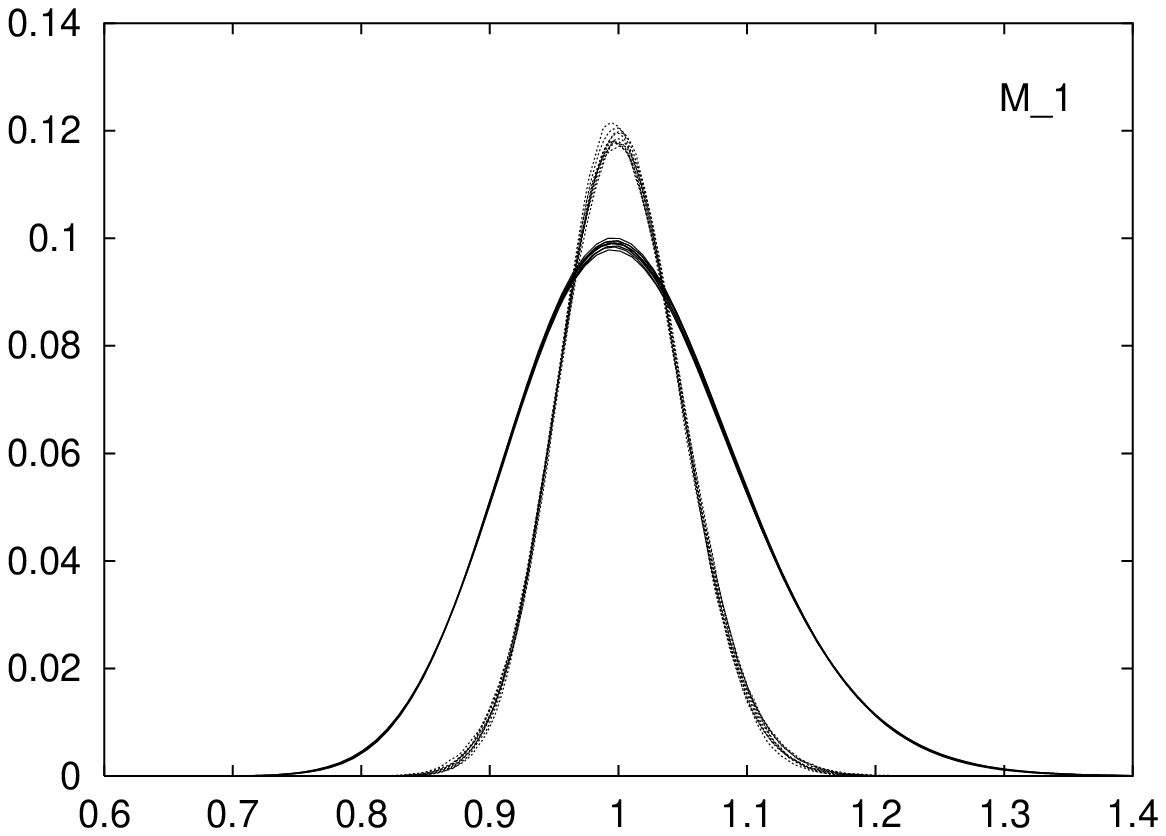,width=.4\hsize}}
  \centerline{\psfig{file=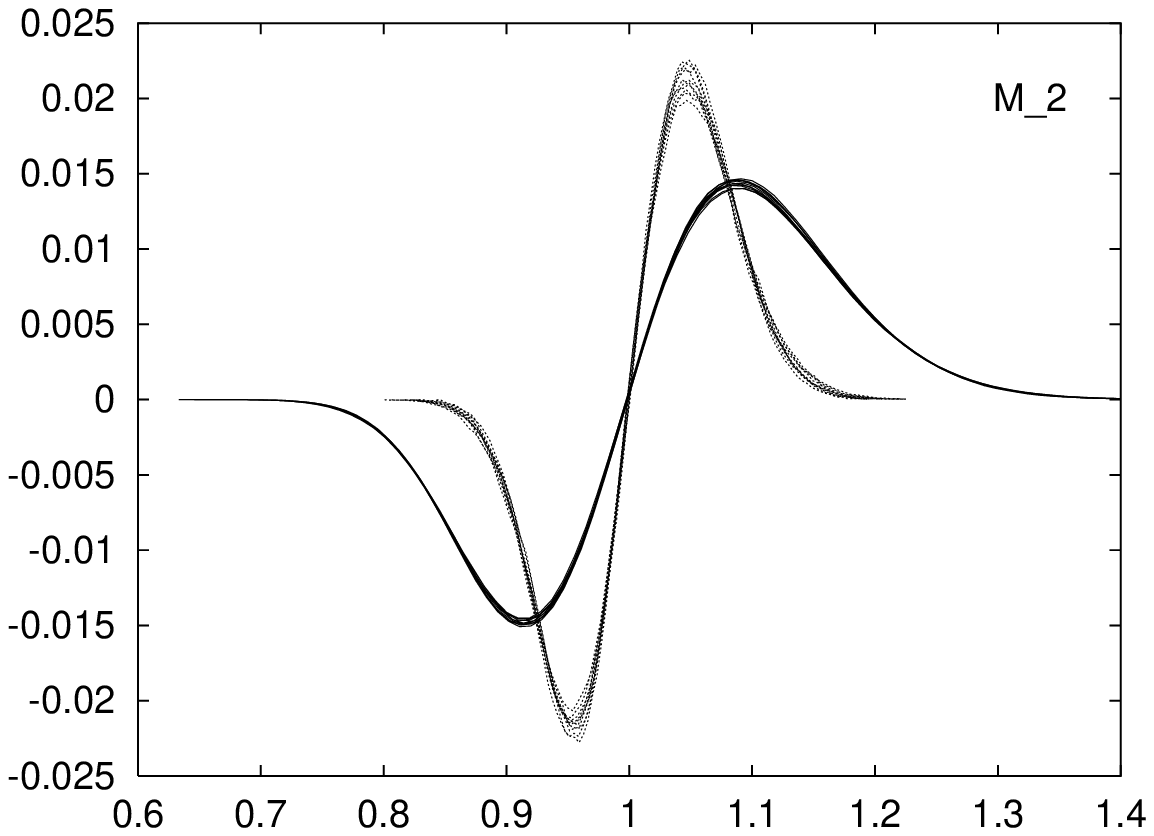,width=.4\hsize}\psfig{file=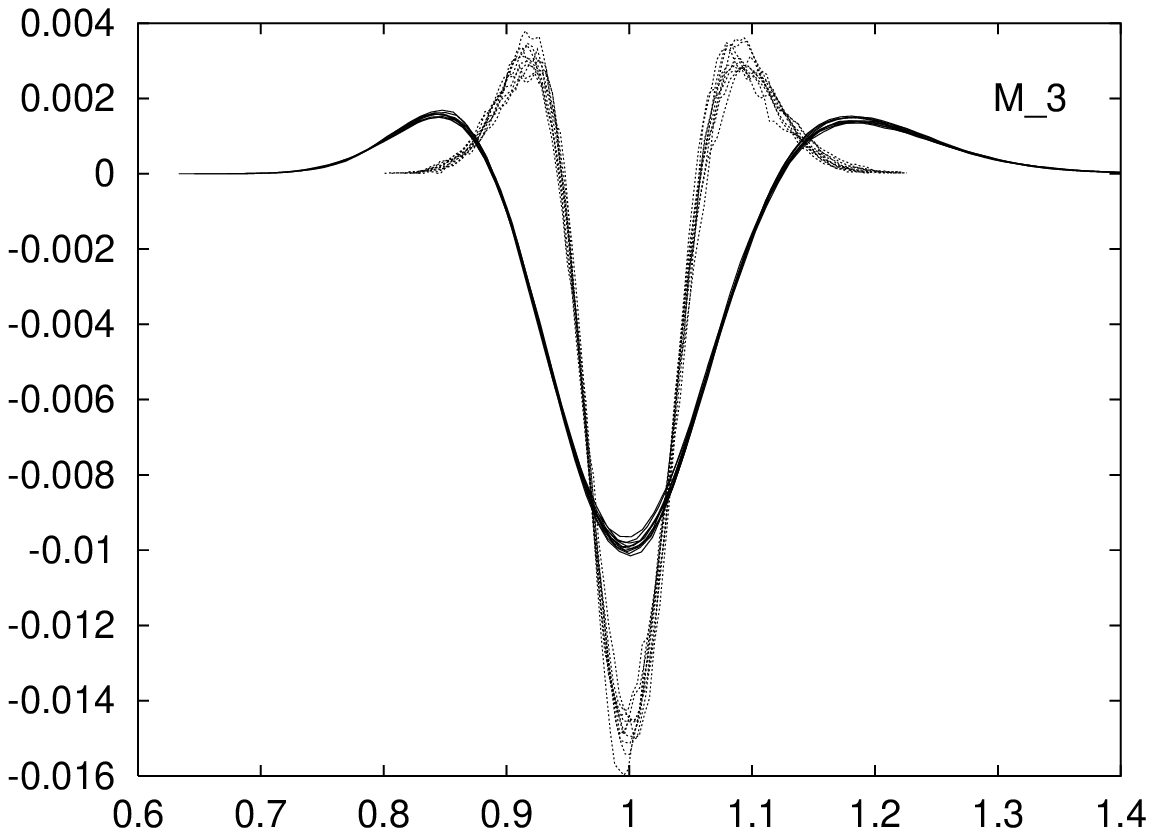,width=.4\hsize}}
    \caption{
      MFs of the ten realizations of the initial matter distribution
      ($z=50$) as a function of the density threshold $1+\delta$, at
      two different spatial resolutions: $128^3$-cell grid (solid
      lines) and $32^3$-cell grid (dotted lines).}
    \label{fig:MFinit}
\end{figure}

Figure~\ref{fig:MFinit} shows the initial MFs as a function of the
density threshold. The general shape of the plots can be explained
qualitatively: for low thresholds, there is no bounding surface at all
(due to the periodic boundary conditions), and hence $M_0 = \mbox{}$
total simulation volume, while $M_1=M_2=M_3=0$.  As the threshold
increases, there appear first disconnected blobs of low density
regions ($M_3 >0$, increasing $M_1>0$, and the boundary surface is
predominantly concave: $M_2<0$), later the blobs fuse together and
tunnels arise ($M_3<0$), and finally the situation reverses and one
ends up with independent blobs of high density regions ($M_3 >0$,
decreasing $M_1>0$, and the boundary surface is predominantly convex:
$M_2>0$), until the threshold becomes larger than the maximum density
($M_0=M_1=M_2=M_3=0$). There are evidences that the zeros of $M_3
(\delta)$ are strongly correlated with the percolation thresholds of
the regions above or below the density threshold (Mecke~\& Wagner
1991).

At the initial time, density fluctuations are small and Gaussian,
which explains the symmetry of the MFs with respect to $\delta=0$.
However, a slight asymmetry can be detected for $M_2$ and $M_3$ on the
$128^3$-cell grid: the spatial resolution is large enough that the MFs
are sensitive to the finite-mass effects induced by the underlying
point-particle distribution. Another difference between the two grids
is the dispersion among realizations, which is larger when the spatial
resolution is small. The scatter in the positions of the zeros and the
values of the extrema tends to increase from $M_0$ toward $M_3$; in
the worst case, the value of the maxima of $M_3(\delta)$ has a
1-$\sigma$ error $\approx 10\%$.

\begin{figure}
     \centerline{\psfig{file=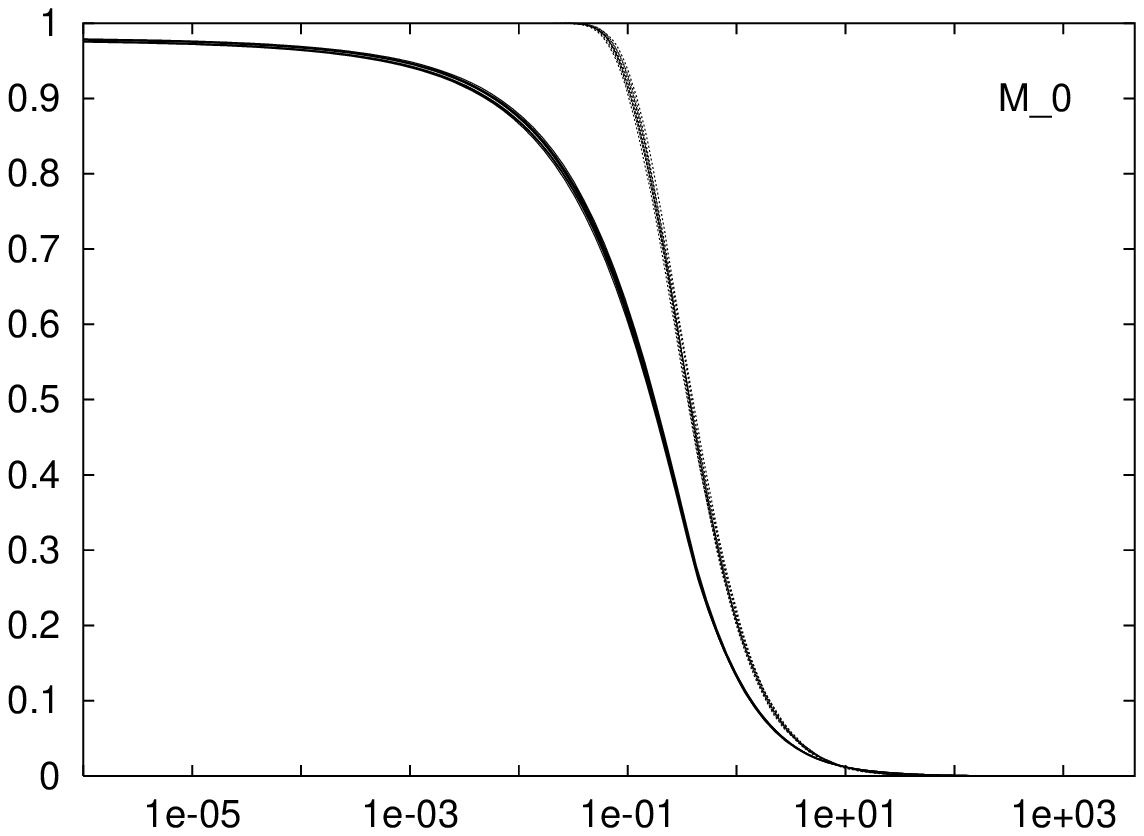,width=.4\hsize}\psfig{file=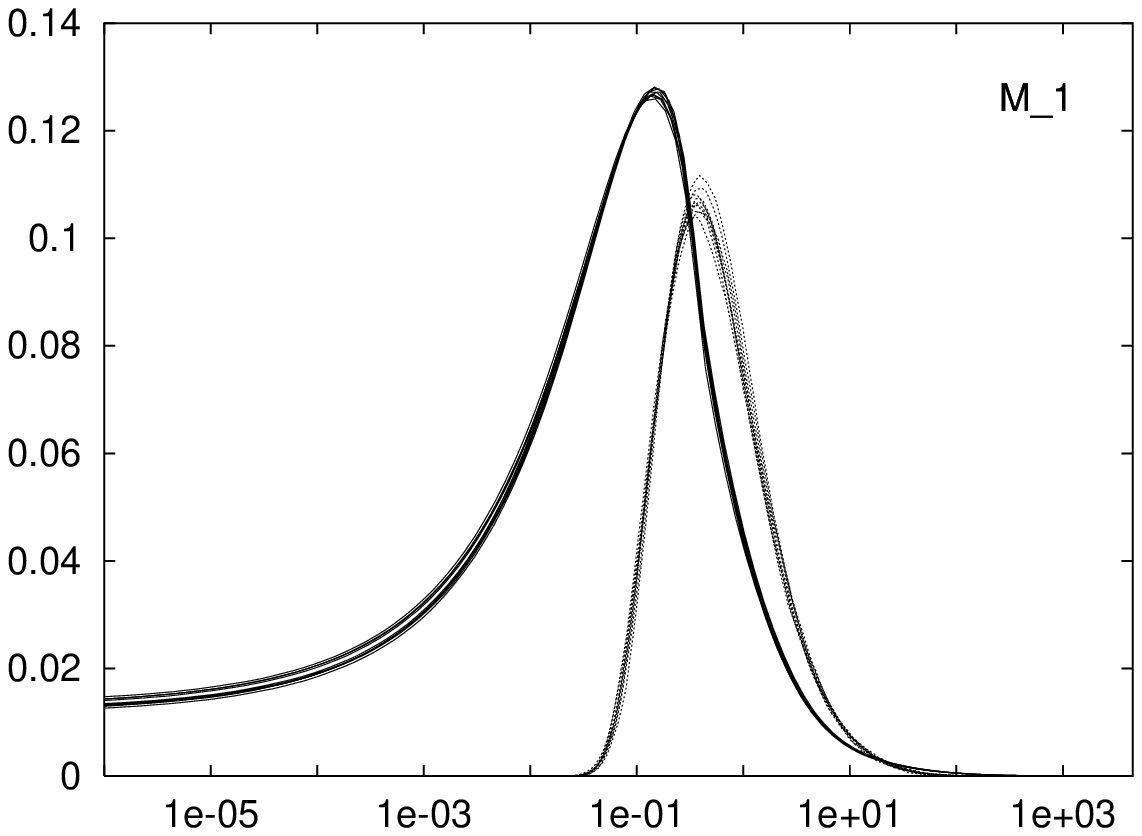,width=.4\hsize}}
     \centerline{\psfig{file=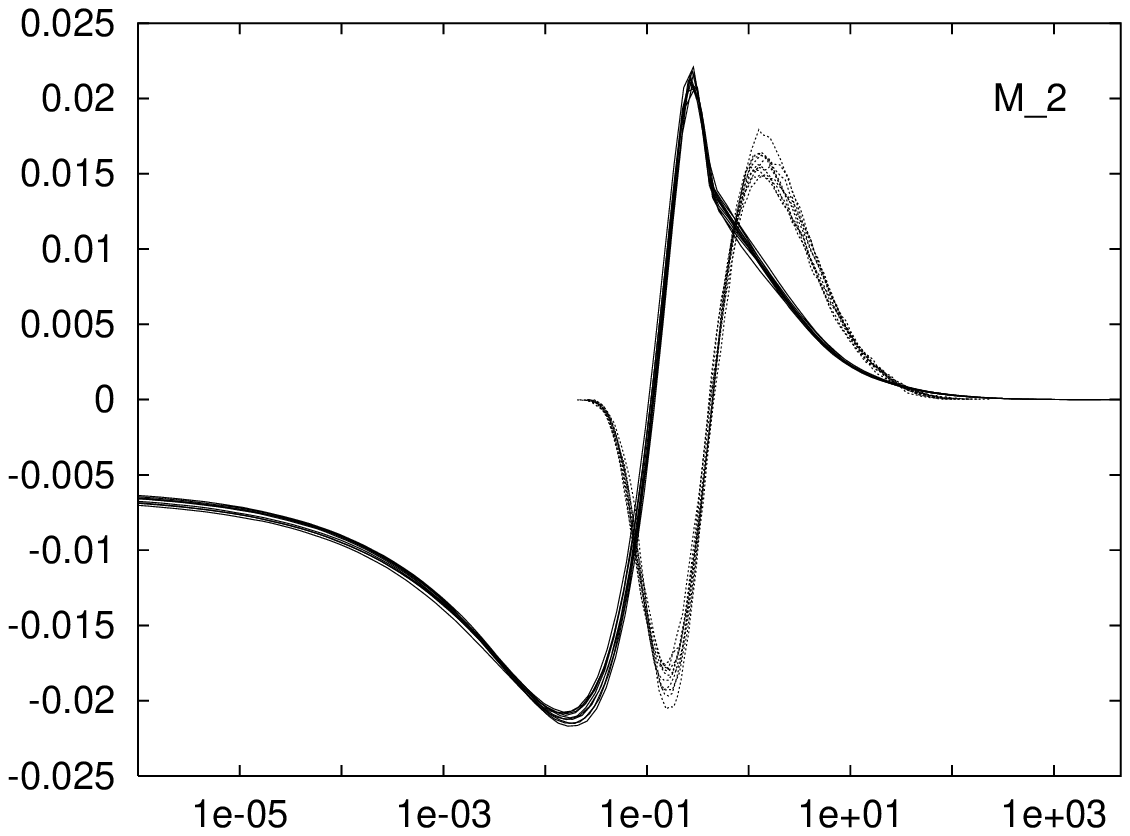,width=.4\hsize}\psfig{file=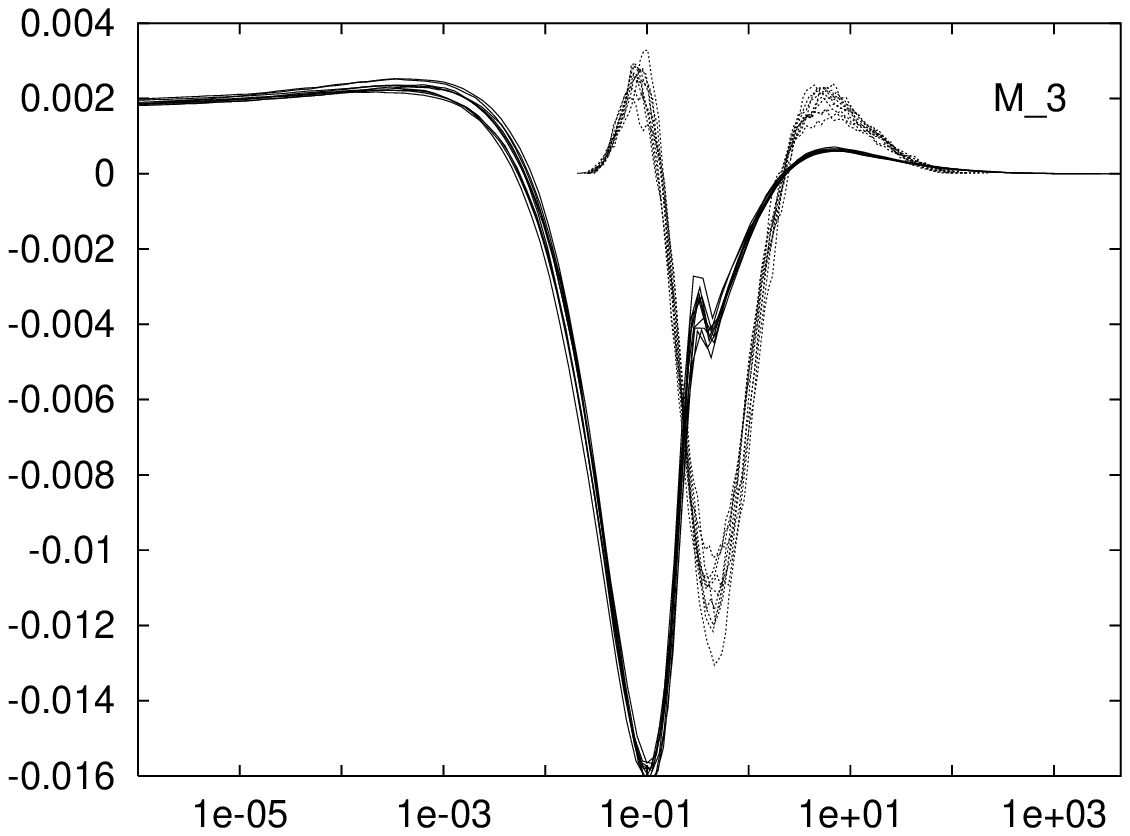,width=.4\hsize}}
    \caption{
      MFs of the ten realizations of the final matter distribution
      ($z=0$) as a function of the logarithm of the density threshold
      $\log (1+\delta)$, at two different spatial resolutions:
      $128^3$-cell grid (solid lines) and $32^3$-cell grid (dotted
      lines).}
    \label{fig:MFfinal}
\end{figure}

Figure~\ref{fig:MFfinal} shows the MFs for the final matter
distribution. The minimum density that can be resolved is
$1+\delta_{min} \approx $ (mean interparticle distance/grid
constant)$^3$, which is $1$ for the $128^3$-cell grid and $1/64$ for
the $32^3$-cell grid. Thus, the curves below these densities are in
principle not reliable\footnote{For instance, the feature at $1+\delta
  \approx 0.4$ of $M_2$ and $M_3$ in the $128^3$-grid is likely a
  finite-mass effect due to isolated particles (density peaks at a
  density $\approx 1$) in the voids of the structure.}. Apart from the
asymmetry around $\delta=0$, one observes in general that the scatter
in the ordinate direction barely changes: the value of the maximum of
$M_3(\delta)$ has an error $\approx 15\%$.
The abscissa-dispersion, however, is larger than at the initial time:
so, e.g., the zeros of $M_3 (\delta)$ in the $32^3$-grid have an error
$\approx 6\%$, while the uncertainty at the initial time is just
$\approx 0.3\%$.

\section{Analysis II: Dark Matter Halos} \label{HaloAnalysis}
The remaining analysis is going to focus on gravitationally bound
halos, identified using the Bound-Density-Maxima method (BDM,
Klypin~\& Holtzman 1997).  We investigate the
scatter in (large-scale) clustering patterns as well as internal
properties of halos introduced by the random nature of the initial
conditions.

\subsection{Identifying Halos} 
We restricted our analysis to halos with more than 100 particles
($M_{\rm vir, min}>10^{12}$\hMsun). This lower mass limit
can be used to derive a lower limit for the virial radius $R_{\rm vir,
min}$ via

\begin{equation}
 M_{\rm vir} = \frac{4\pi}{3} \Delta_{\rm vir} \rho_b R_{\rm vir}^3 \ ,
\end{equation}

\noindent
where $\rho_b$ is the background density and $\Delta_{\rm vir} = 340$
for the \LCDM\ model under consideration. 

The BDM code identifies local overdensity peaks by
smoothing the density field on a particular scale. 
The particle
distribution was used to iteratively find potential halo centers as the centers of mass of 20,000 spheres
with radius $R_{\rm sphere} \approx R_{\rm vir, min} \approx
162$\hkpc\ centered about randomly chosen particles. Once the iteration
converged for all spheres we repeated the procedure using successively
smaller sphere radii down to 70\hkpc, about three
times the force resolution. For each of these halo centers we stepped
out in radial bins until the density drops below $\rho_{\rm bin} <
\Delta_{\rm vir} \rho_b$. This defined the outer radius $R_{\rm vir}$
of the halo\footnote{If we want to identify halos-within-halos this
method needs to be adjusted to account for the fact that the actual
density of a satellite galaxy might not drop below $\Delta_{\rm vir}
\rho_b$}. We discarded all halos with less than 100 particles within
$R_{\rm vir}$ for the further analysis.\footnote{To cross check the
completeness of our BDM halo catalogues we also performed a FOF
analysis which shows a nearly 100\% agreement and only an
incompleteness in the BDM catalogues for halos less massive than 100
particles.}

\subsection{Mass Function of Halos} 

   \begin{figure}
      \centerline{\psfig{file=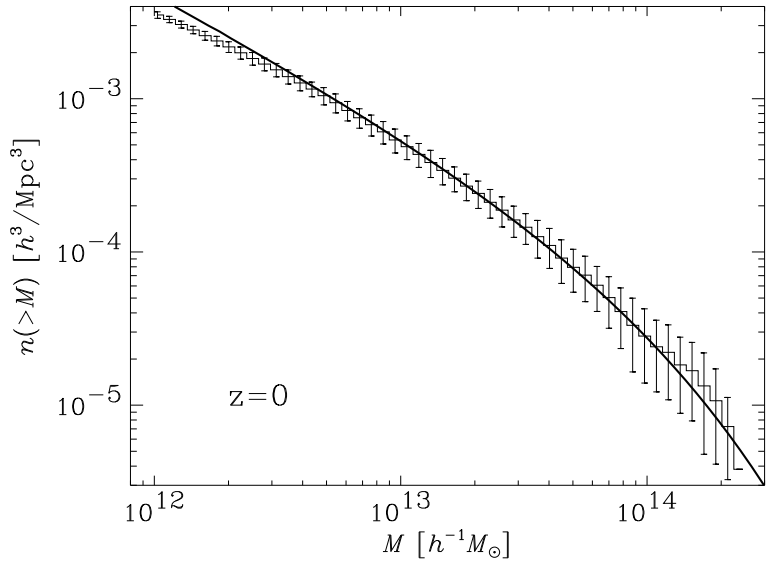,width=11cm}}
      \caption{Cumulative mass functions of BDM halos compared
               to Press-Schechter prediction (Press~\& Schechter 1974).
               The mass functions is the average taken over all 
               ten runs and the error bars are $1\sigma$ errors.}
      \label{Massfunc} 
   \end{figure}
 
The first quantity to investigate is the mass spectrum.  We calculated
the cumulative mass function $n(>\!M)$  for our BDM halos and compared it to the
analytical prediction of Press~\& Schechter (1974):

\begin{equation} \label{PS}
 \displaystyle \frac{dn}{dM} dM = 
 \sqrt{\frac{2}{\pi}} \frac{\langle \rho\rangle}{M} \frac{\delta_c}{\sigma_M}
                    \left|\frac{d \ln \sigma_M}{d \ln M}\right|
                    \exp{\left(-\frac{\delta_c^2}{2 \sigma_M^2}\right)}
                    \frac{dM}{M} ,
\end{equation}

\noindent
where the variance $\sigma_M$ is given by \Eq{sigmarAna} and
\mbox{$\delta_c$ = 1.68}.

\Fig{Massfunc} shows that the average mass function of all ten runs
is in good agreement with the PS prediction, what has been noted
already by several other authors (Efstathiou~\ea 1988; White,
Efstathiou~\& Frenk 1993; Gross~\ea 1998; Governato~\ea 1999;
Jenkins~\ea 2001). This again is another indicator that the initial
conditions as well as the evolution by \nbody\ simulation are in fair
agreement with theoretical predictions based on the analytical power
spectrum and its time evolution. The discrepancy of the numerical
$n(>\!M)$ with the PS prediction at the low and high mass end of the
mass function is also a well known fact (e.g. Governato~\ea 1999) and
not related to unreliable ICs or wrong \nbody\ modeling.  Anyway, we
are more interested in the scatter stemming from the random nature of
the initial conditions.  We are driven by the question, to what extent
a single cosmological simulation can be representative for the volume
under investigation.  We observe that the scatter gradually increases
from around 4\% at the very low mass end resolved to about 50\% for
the most massive objects found in the simulation.
According to the PS prediction, the scatter due to cosmic variance
should enter via the amplitude $A_{\vec{k}}$ predominantly, not the
phases $\theta_{\vec{k}}$ of the ICs, Eq.~(\ref{Pkcomplex}). The
observed increase of scatter with mass is then naturally explained
also by the PS formula, given that larger masses are more sensitive to
larger scales.

\subsection{Halo-Halo Correlation Function}
   \begin{figure}
      \centerline{\psfig{file=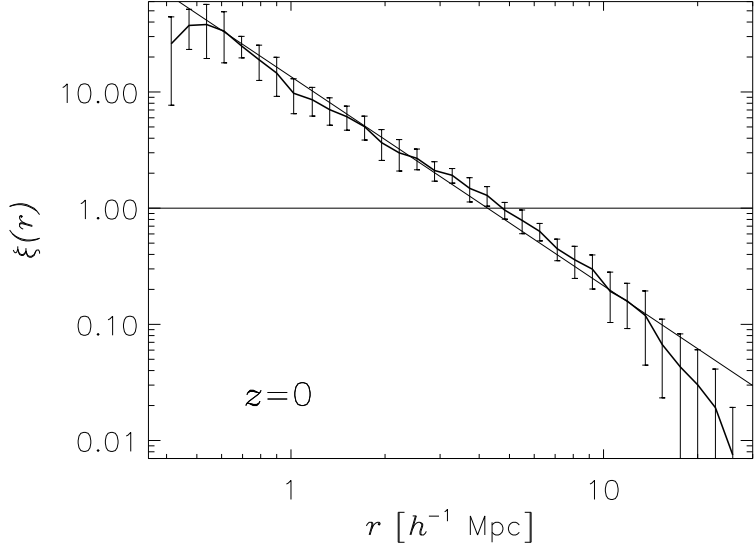,width=11cm}}
      \caption{Two-point correlation function $\xi(r)$ for BDM halos at
               redshift $z=0$. Error bars are again 1$\sigma$.
               The thin solid line is the fit to a power law as
               given by \Eq{xi}.}
      \label{XiR-halo} 
   \end{figure}

The calculation of the halo-halo two-point correlation function is
based on the estimator \Eq{xirSimu} again.  However, this time we
applied it only to the 500 most massive objects in the runs, which means
fixing the number density of halos to $n_{\rm halo}=2 \cdot
10^{-3}$(\hMpc)$^{-3}$. This choice for $n_{\rm halo}$ restricts the
masses of the objects used from $M \sim 3 \cdot 10^{14}$\hMsun\ down
to $M \sim 2 \cdot 10^{12}$\hMsun. The result for the average taken
over the ten BDM catalogs at redshift $z=0$ is shown in
\Fig{XiR-halo}. The mean correlation function $\xi_{\rm est}(r)$ was 
 fitted to a power law,

\begin{equation} \label{xi}
 \xi(r) = (r_0/r)^\gamma \ ,
\end{equation}

\noindent
over the range $r\in[0.5,20]$\hMpc\ with the parameters $r_0$=4.26
$\pm0.44$\hMpc\ and $\gamma$=1.80 $\pm0.17$. The 1$\sigma$ errors are
of the order of 10\% and indicate again only a mild dependence of the
halo-halo correlation function on the variance introduced by the
random nature of the initial conditions. Even though the scatter for
the fundamental mode is $\approx 20\%$, it does only marginally affect
the statistical clustering properties of dark matter halos in the
respective mass range.

\subsection{Internal Properties of the Most Massive Halo} 

   \begin{figure}
      \centerline{\psfig{file=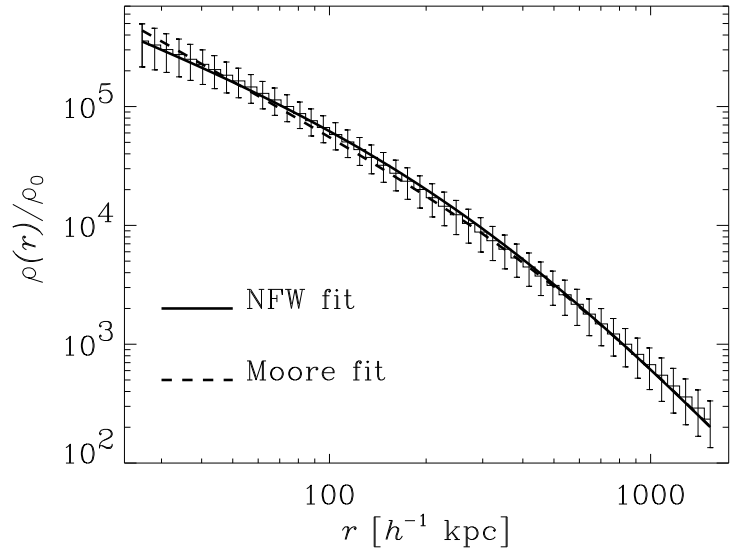,width=11cm}}
      \caption{Average density profile for the most massive BDM halo
               with 1$\sigma$ error bars along with a NFW and
               a Moore profile fit to the data.}
      \label{DensProfile} 
   \end{figure}

Even though we are only resolving approximately 2\% of the virial
radius of the most massive particle group (cf. \Table{properties}), we
fitted a Navarro, Frenk~\& White (NFW) profile (Navarro, Frenk~\&
White 1997),

\begin{equation}\label{NFWfit}
 \rho_{\rm NFW}(r) \propto \frac{1}{r/r_s (1+r/r_s)^2} ,
\end{equation}

\noindent
as well as a Moore profile (Moore~\ea 1999), 

\begin{equation} \label{MooreFit}
 \rho_{\rm Moore}(r) \propto \frac{1}{(r/r_s)^{1.5} (1+(r/r_s)^{1.5})} ,
\end{equation}

\noindent
to the most massive halo found in the BDM catalogues. The
question we are interested in is whether the scatter due to the random
nature of the initial conditions can be made responsible for the
difference in the central slope of the density profile described by
those two fitting formulae. And from
\Fig{DensProfile} we deduce that at least down to the resolved scale
of 2\% of the virial radius both analytical descriptions for the
density profile do give indistinguishable good fits to the actual
data; they both lie within the 1$\sigma$ error bars. However, we must
stress that both profiles start to deviate even stronger from each
other for even smaller scales not covered by the current
study. Moreover, the reduced $\chi^2$ value for the NFW fit is
marginally better than for the Moore fit, as one might anticipate from
the behavior at small $r$ values in
\Fig{DensProfile}.

We conclude the analysis with \Table{properties} summarizing
some internal properties calculated for the most massive halo, i.e.
mass $M$, circular velocity $v_{\rm circ}$, velocity dispersion
$\sigma_v$, virial radius $r_{\rm vir}$, concentration parameter

\begin{equation} \label{concentration}
 c = r_{\rm vir} / r_s  ,
\end{equation}

\noindent
where $r_s$ is the scale radius derived from the fit to the NFW
profile~\Eq{NFWfit}, the spin parameter

\begin{equation} \label{xlambda}
 \lambda = J \sqrt{|E|} / (G M^{5/2})  ,
\end{equation}

\noindent
and the triaxiality parameter

\begin{equation} \label{triaxial}
 T = \displaystyle \frac{a^2 - b^2}{a^2 - c^2} \ ,
\end{equation}

\noindent
where $a>b>c$ are the eigenvalues of the inertia tensor.

\begin{table}
\caption{Internal (averaged) properties of the most massive halo when
         averaged over ten runs. The errors are the 1$\sigma$-value
         again.}
\label{properties}
\begin{center}
\begin{tabular}{llllc}

\multicolumn{4}{l}{property} & variance \\ \hline \hline
\\
$M$            & = (3.07  & $\pm$ 1.60)  & $10^{14}$\hMsun   & 52 \%\\
$v_{\rm circ}$ & = (1131  & $\pm$ 199)   & km/sec            & 18 \%\\
$\sigma_v$     & = (1172  & $\pm$ 195)   & km/sec            & 17 \%\\
$r_{\rm vir}$  & = (1344  & $\pm$ 163)   & \hkpc             & 12 \%\\
$c$            & = \ 4.10  & $\pm$ 0.91  &                   & 22 \%\\
$\lambda$      & = \ 0.033 & $\pm$ 0.018 &                   & 53 \%\\
$T$            & = \ 0.762 & $\pm$ 0.102 &                   & 13 \%\\

\end{tabular}
\end{center}
\end{table}

This Table shows that the 1$\sigma$ variance for nearly all quantities
is of the order of 20\%, like the variance of the fundamental mode
$k_{\rm min}=2\pi/B$ (cf. \Fig{Power64xx}). Only the mass and the spin
parameter show a larger scatter.

\section{Summary and Conclusions}
We presented the study of ten random realizations of a density field
characterized by a cosmological power spectrum $P(k)$ at redshift
$z=50$. These initial conditions for \nbody\ simulations were tested
with respect to their correlation properties. Recent claims by
Baertschiger~\& Sylos Labini (2002) throw doubts on the ability of the
commonly used method for generating the initial density field using
particles (i.e. glass or grid preinitial distribution + the Zeldovich
approximation, Eftstahiou~\ea 1985) to clearly reproduce the
analytical input correlations. 
The power spectrum $P(k)$ and the mass variance $\sigma_M(r)$ do not
deviate from the expected behavior (including expected departures from
the desired \LCDM\ behavior due to finite mass and finite size
effects). The estimated 2-point correlation $\xi(r)$ is too noisy
to be used as a reliable credibility check; one cannot claim either that it reproduces the desired \LCDM\ 
behavior or that it exhibits systematic deviations thereof.
%

These initial conditions were then evolved forward in time until
redshift $z=0$ using the publicly available adaptive mesh refinement
code \mlapm\ (Knebe, Green~\& Binney 2001). This allowed us to explore the cosmic variance
stemming from the random nature of the initial conditions,
i.e., the scatter between different realizations of statistically
identical initial conditions.  
We addressed the morphological properties of the matter distribution
with the four Minkowski functionals as functions of a density
threshold. The scatter grows in time, the one exhibiting a larger
dispersion being the genus, of the order of $10\%$ at $z=0$.  We also
investigated the internal properties of DM halos, which have already
been shown by other groups to be profoundly influenced by the
surrounding large-scale structure, which in turn is sensitive to
$k$-modes $\approx$ fundamental mode (Colberg~\ea 1999).
We find that the scatter in the properties of the most massive
object(s) forming in the box is $\sim 20\%$, and as high as $\sim
50\%$ for some properties such as the mass or the spin parameter.



An interesting question is whether this scatter is
induced mainly by the cosmic variance of the amplitude at scales
around the fundamental mode, or by the cosmic variance of the
random phases. There is certainly a propagation of the error in the
initial large-scale amplitude by power transfer towards smaller
scales. In fact,
when comparing our data to the (non-)linear fit of Peacock~\& Dodds
(1996) for the power spectrum and to the prediction by Press~\&
Schechter (1974) for the mass function, we find good agreement. The
derivation of both results is based on the hypothesis of small
influence from coupling of modes at some $k$ to modes with larger $k$; our results support
this assumption, as far as the statistical estimators we probed are
concerned. It would now be interesting to investigate in detail the
actual influence of the large waves onto the small scale
structure. This would also shed some light on the credibility of
running small simulation boxes to very low redshifts as already done
by several groups (e.g. Dave~\ea 2001, Avila-Reese~\ea 2001, Gnedin
2000, Colin, Avila-Reese~\& Valenzuela 2000), but we leave this to a
future study.

\section*{Acknowledgments}
We would like to thank Brad Gibson for a careful reading of the
manuscript and valuable comments. AK greatly acknowledges the
hospitality of Rosa Dom\'\i nguez Tenreiro and Gustavo Yepes at
Universidad Aut\'onoma de Madrid where this work was started. AD
thanks K. Mecke for the code to compute the Minkowski
functionals.

The simulations presented in this paper were carried out on the
Beowulf cluster at the Centre for Astrophysics \& Supercomputing,
Swinburne University. AK acknowledges the support of the Swinburne
University Research Development Grants Scheme.

\section*{References}

\reference
        {Avila-Reese V., Colin P., Valenzuela O., D'Onghia E.,
         Firmani C., \ApJ{559}{516}{2001}}

\reference
        {Baertschiger T., Sylos Labini F., Europhys. Lett. {\bf 57} 322 (2002)}

\reference
        {Colberg J.M., White S.D.M., Jenkins A., Pearce F., \MNRAS{308}{593}{1999}}

\reference
        {Colin P., Avila-Reese V., Valenzuela O., \ApJ{542}{622}{2000}}

\reference
        {Dave R., Spergel D.N., Steinhardt P.J., Wandelt B.D.,
         \ApJ{547}{574}{2001}}

\reference
        {Efstathiou G., Frenk C.S., White S.D.M., \ApJS{57}{241}{1985}}

\reference
        {Efstathiou G., Frenk C.S., White S.D.M., Davis M., \MNRAS{235}{715}{1988}}

\reference
        {Gabrielli A., Joyce M., Sylos Labini F., \PhRevD{65}{083523}{2002}}

\reference
        {Gnedin N.Y., \ApJ{542}{535}{2000}}

\reference
        {Governato F., Babul A., Quinn T., Tozzi P., Baugh C.M., 
         Katz N., Lake G., \\
         \MNRAS{307}{949}{1999}}

\reference
        {Gross M.A.K., Somerville R.S., Primack J.R., Holtzmann J.,
         Klypin A.A., \\
         \MNRAS{301}{81}{1998}}

\reference
        {Hamana T., Yoshida N., Suto Y., \astroph{0111158}}

\reference
        {Jenkins A., Frenk C.S., Pearce F.R., Thomas P.A., Colberg J.M.,
         White S.D.M., Couchman H.M.P., Peacock J.A., Efstathiou G.,
         Nelson A.H., \ApJ{499}{20}{1998}}

\reference
        {Jenkins A., Frenk C.S., White S.D.M., Colberg J.M.,
         Cole S., Evrard A.E., Couchman H.M.P., Yoshida N., 
         \MNRAS{321}{372}{2001}}

\reference
{Kerscher M., Schmalzing J., Retzlaff J., Borgani S., Buchert T., 
  Gottloeber S., Mueller V., Plionis M., Wagner H., \MNRAS{284}{73}{1997}

\reference
        {Klypin A.A., Holtzman J., \astroph{9712217}}

\reference
        {Knebe A., Green A., Binney J., \MNRAS{325}{845}{2001}}

\reference
{Mecke K.R., Buchert T., Wagner H., \AaA{288}{697}{1994}}

\reference
{Mecke K.R., Wagner H., J. Stat. Phys. {\bf 64}, 843 (1991)}

\reference
{Melott A. L., Phys. Rep. {\bf 193}, 1 (1990)}

\reference
        {Moore B., Quinn T., Governato F., Stadel J., Lake G.,
         \MNRAS{310}{1147}{1999}}

\reference
        {Navarro J., Frenk C.S., White S.D.M., \ApJ{490}{493}{1997}}

\reference
        {Peacock J.A., Dodds S.J., \MNRAS{280}{L19}{1996}}

\reference
        {Pen U.L., \ApJL{490}{127}{1997}}

\reference
        {Press W.H., Schechter P., \ApJ{187}{425}{1974}}

\reference
        {Press W.H., Teukolsky S.A., Vetterling W.T., Flannery B.P., in
        {\em Numerical Recipes}, Cambridge Univ. Press, 1992}

\reference
{Schmalzing J., Buchert T., Melott A.L., Sahni V., Sathyaprakash B.S., Shandarin S.F., \ApJ{526}{568}{1999}

\reference
{Schmalzing J., Gorski  K. M., \MNRAS{297}{355}{1998}

\reference
        {Seljak U., Zaldarriaga M., ApJ {\bf 469}, 437 (1996)}


\reference
        {White S.D.M., in \textit{Cosmology and Large-Scale Structure}, 
         eds. Schaeffer R., Silk J., Spiro M., Zinn-Justin J., Elsevier 1996, 
         p. 349}

\reference
        {White S.D.M., Efstathiou G., Frenk C.S., \MNRAS{262}{1023}{1993}}


\end{document}